\documentclass[osajnl2,twocolumn,showpacs]{revtex4}
\usepackage{here}
\usepackage{graphicx}
\graphicspath{{pict/}{}}

\usepackage{bm}

\newcounter{Fig}

\newcommand{\be}{\begin{equation}}
\newcommand{\ee}{\end{equation}}

\begin{document}

\title{Interface solitons in two-dimensional photonic lattices}

\author{Mario I. Molina}

\affiliation{Departmento de F\'{\i}sica, Facultad de Ciencias,
Universidad de Chile, Casilla 653, Santiago, Chile}

\author{Yuri S. Kivshar}

\address{Nonlinear Physics Center, Research School of Physical Sciences
and Engineering, Australian National University, Canberra ACT 0200,
Australia}


\begin{abstract}
We analyze localization of light at the interface separating
square and hexagonal photonic lattices, as recently realized
experimentally in two-dimensional laser-written waveguide  arrays 
in silica glass with self-focusing nonlinearity [A. Szameit {\em
et al.}, Opt. Lett. {\bf 33}, 663 (2008)]. We reveal the
conditions for the existence of {\em linear} and {\em nonlinear}
surface states substantially influenced by the lattice topology,
and study the effect of the different symmetries and couplings on
the stability of two-dimensional interface solitons.
\end{abstract}


\maketitle

Theoretical results on the existence of novel types of discrete
surface solitons localized in the corners or at the edges of
two-dimensional photonic lattices~\cite{makris_2D,pla_our,pre_2D}
have been recently confirmed by the experimental observation of
two-dimensional surface solitons in optically-induced photonic
lattices~\cite{prl_1} and two-dimensional waveguide arrays
laser-written in fused silica~\cite{prl_2,ol_szameit}. These
two-dimensional nonlinear surface modes demonstrate novel features
in comparison with their counterparts in truncated one-dimensional
waveguide arrays~\cite{OL_george,PRL_george,OL_molina}. In
particular, in a sharp contrast to one-dimensional discrete
surface solitons, the mode threshold is lower at the surface than
in a bulk making the mode excitation easier~\cite{pla_our}.

Recently, Szameit {\em et al.}~\cite{interface_2D} reported on the
first experimental observation of two-dimensional interface
solitons, i.e. spatial optical solitons generated at the interface
separating square and hexagonal optical lattices with different
refractive index modulation depths. Such two-dimensional interface
solitons feature {\em asymmetric shapes}, while differences in
array properties and lattice topology strongly affect the
threshold power for soliton existence and excitation as well as
soliton stability.

In this Letter, we study this problem analytically in a more
general setting and analyze localization of light at the interface
separating two lattices of different symmetries in the framework
of the two-dimensional discrete nonlinear model. We assume that
the interface is created between the square and hexagonal
two-dimensional optical lattices, similar to the case studied
earlier~\cite{interface_2D}, but we assume the coupling parameters
to be different and study the effect of different symmetries and
lattice topology, as well as the coupling strength of the
interface on the existence and stability of both {\em linear} and
{\em nonlinear} surface states. In particular, we determine the
conditions for the thresholdless surface states (linear surface
modes) that appear due to the breaking of the lattice topology,
and also study the nonlinear localization and generation 
\begin{figure}[t]
\noindent\includegraphics[scale=0.35,angle=0]{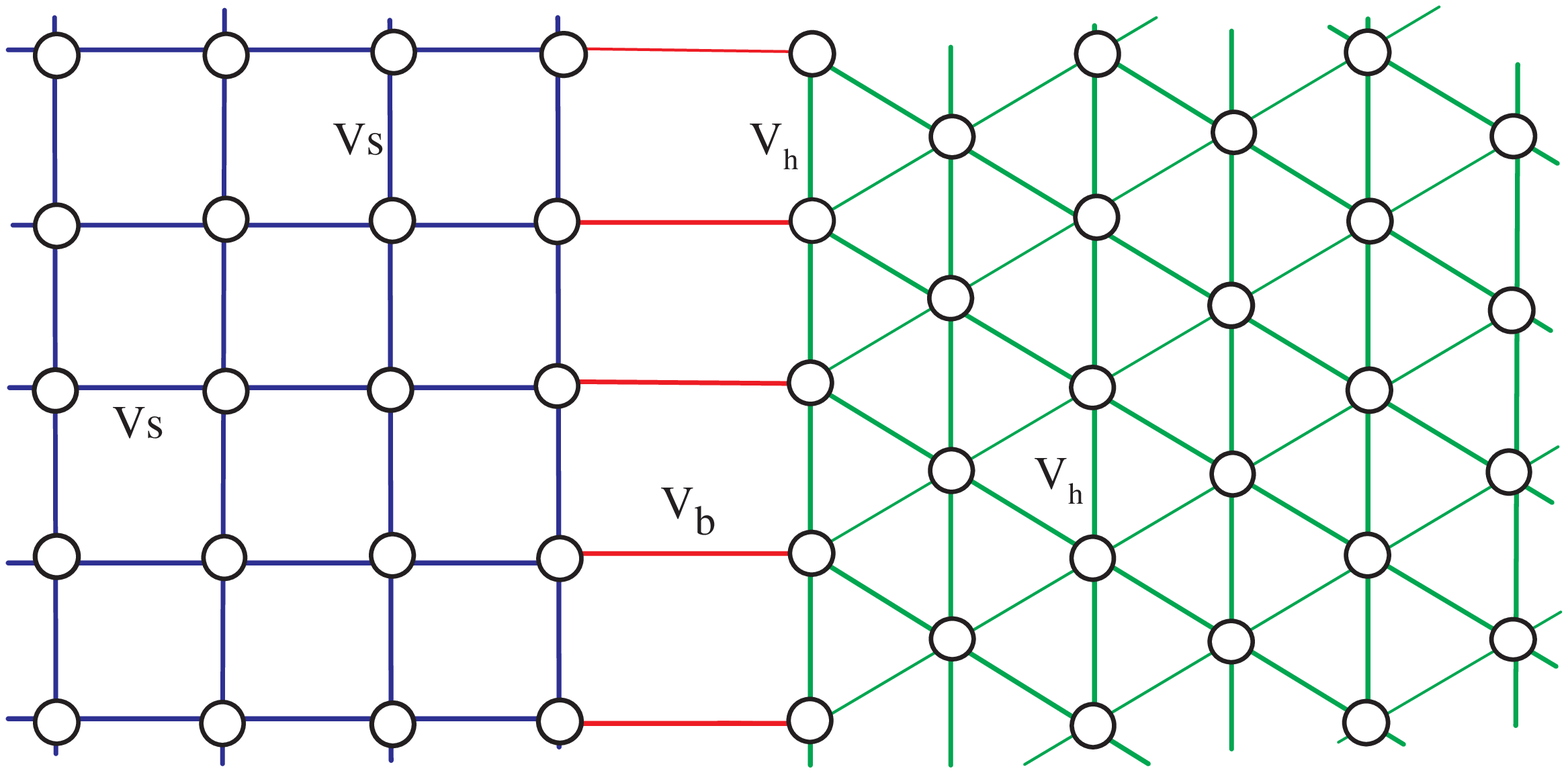}\hspace{6.0 cm}
\includegraphics[scale=0.46,angle=0]{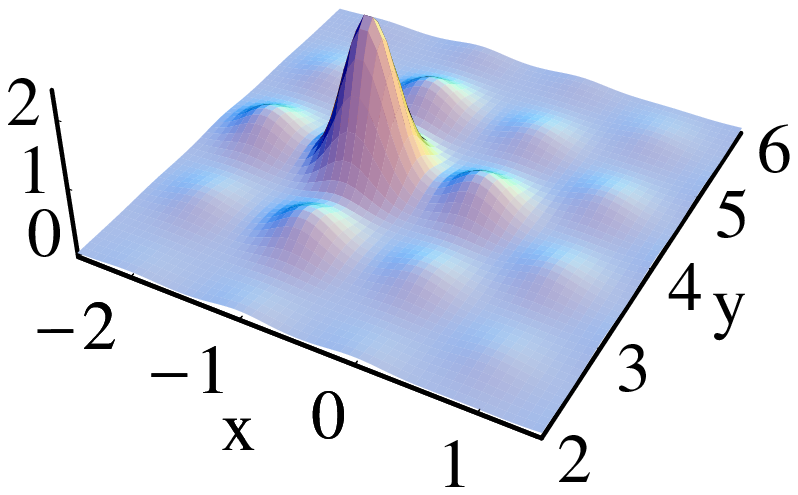}
\includegraphics[scale=0.48,angle=0]{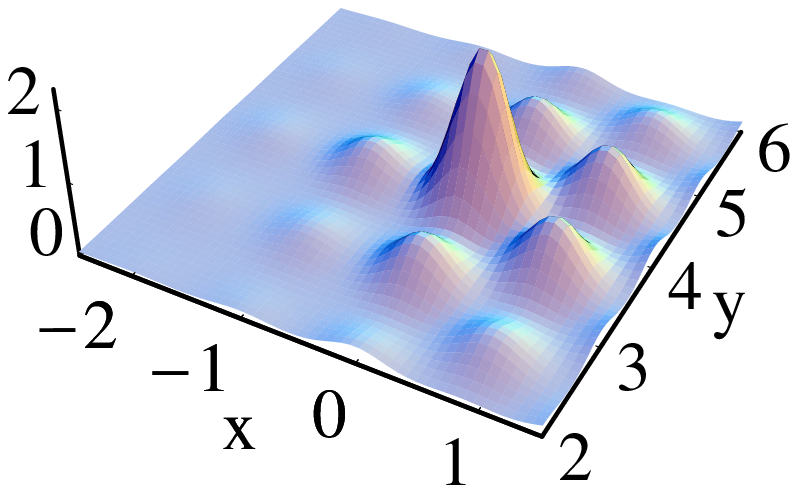}
\caption{(Color online) Top: Sketch of a square-hexagonal photonic
lattice. Bottom: Examples of low-order nonlinear interface mode
centered (left) on the site belonging to the boundary of the
square lattice (which ends at $x=-1$), and (right) on the site
belonging to the boundary of the hexagonal lattice (which starts
at $x=0$).} \label{fig1}
\end{figure}
of two-dimensional interface solitons.

We consider a semi-infinite two-dimensional optical lattice
created by a square array of optical waveguides joined to the
other semi-infinite but hexagonal lattice along a straight
boundary, as shown in Fig.~1 (top).

In the framework of the coupled-mode theory, the electric field
${\cal E}({\bf r})$ propagating along the waveguides can be presented as
a superposition of the waveguide modes, ${\cal E}({\bf
r})=\sum_{\bf n} {\cal E}_{\bf n} \phi({\bf r}-{\bf n})$, where
${\cal E}_{{\bf n}}$ is the amplitude of the (single) guide mode
$\phi({\bf r})$ centered on site with the
\begin{figure}[t]
\noindent
\includegraphics[scale=0.4,angle=0]{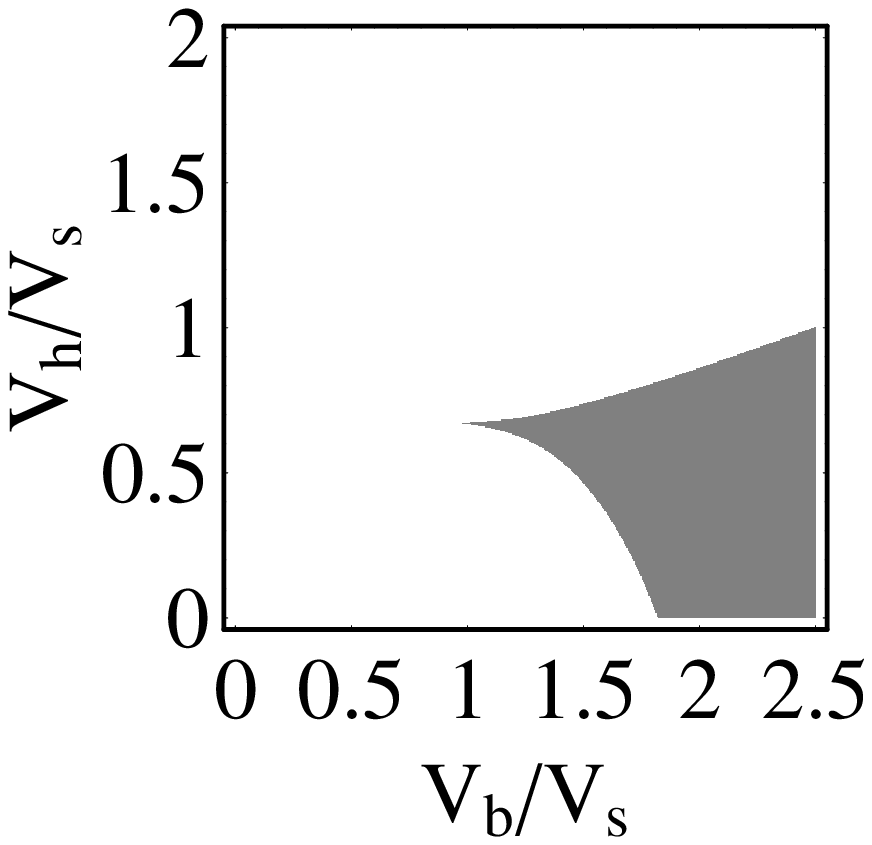}\hspace{0.0cm}
\includegraphics[scale=0.5,angle=0]{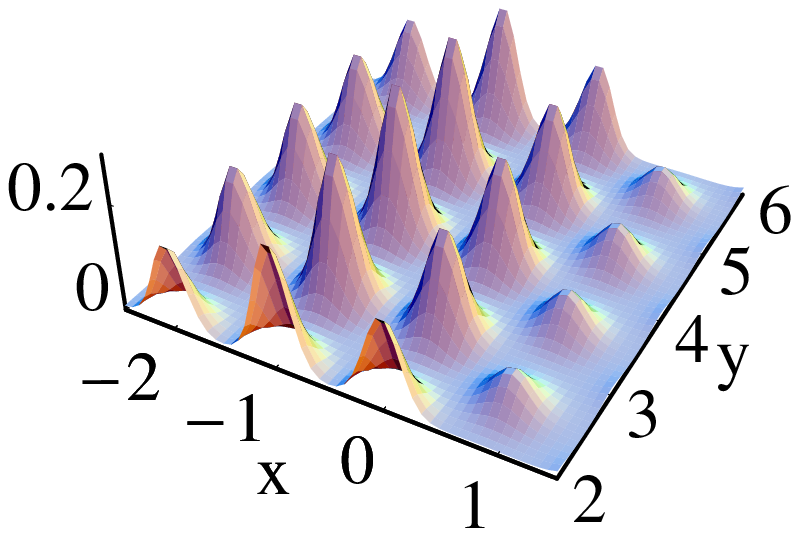}
\caption{(Color online) Linear interface localized modes. (a)
Existence diagram with the shaded area for the modes centered at
the interface. Right: Example of a linear localized mode for
$V_{b}/V_{s}=2,V_{h}/V_{s}=0.5$. Square lattice ends at $x=-1$
while triangular lattice starts at $x=0$.}\label{fig2}
\end{figure}
lattice number ${\bf n}= (n_1,n_2)$.  The evolution equations
for the modal amplitudes $E_{\bf n}$ take the form,
\be i {d {\cal E}_{\bf n}\over{dz}} + V \sum_{n_1,n_2}{\cal
E}_{\bf m} + \gamma |{\cal E}_{\bf n}|^2 {\cal E}_{\bf n} =
0,\label{eq:1}
\ee
where ${\bf n}$ denotes the position of a guide center, and the coupling $V$
takes on the values $V_{s}$( or $V_{h}$) inside
the square (or hexagonal) lattice, and $V_{b}$, along the boundary
between both lattices (see notation in Fig.~1). The nonlinear parameter $\gamma$ is normalized to
$1(-1)$ for the focussing (defocussing) nonlinearity. The lattice studied in
this Letter contains 96 sites with open boundary conditions.

Next, we analyze the stationary localized modes of Eq.(\ref{eq:1})
of the form ${\cal E}_{n}(z) = E_{n} \exp(i \beta z)$, where the
amplitudes $E_{n}$ satisfy the nonlinear difference equations,
\be %
 -\beta E_{n} + V \sum_{n_1,n_2}E_{\bf m} + \gamma |E_{\bf
n}|^2 E_{\bf n} = 0\label{eq:2}
\ee
We start our analysis by studying the fundamental (nodeless) modes
of the system (\ref{eq:2}) in the linear limit (i.e. for
$\gamma=0$) which corresponds to low input powers $P=\sum_{\bf n}
|E_{\bf n}|^2$. After normalizing the lattice couplings to one
corresponding to the square lattice, we are left with two
independent parameters, $V_{h}/V_{s}$ and $V_{b}/V_{s}$. For given
values of these two parameters, we diagonalize the appropriate
matrix and examine its fundamental (nodeless) mode.  In general,
these modes are wide and highly asymmetric in the direction
perpendicular to the interface. The position of the mode center is
very sensitive to the specific value of the coupling parameters.
For instance, for $V_{t}/V_{s}=V_{b}/V_{s}=1$, the mode center is
shifted inside the hexagonal lattice so that the interface acts as
a {\em repulsive potential}, preventing the light beam from
crossing the interface. Thus, we observe that the lattice topology
and geometry are playing an important role. We attribute this
effect to the {\em coordination number mismatch} that can be
easily compensated by changing the interactions between the
waveguides in each lattice. Indeed, if we compensate for this
geometric effect by setting the ratio $V_{h}/V_{s}$ equal to the
ratio of their respective coordination numbers, $2/3$, while
keeping $V_{b}=V_{s}$, {\em no mismatch} is observed and the
resulting mode extends over both lattices.

We now make a sweep in coupling space selecting those values that
lead to a mode center located at the interface either from the
side of the square lattice or from the side of the hexagonal
lattice. Our results are summarize in Fig.~2 in the form of a
coupling-parameters diagram, along with an example of one such
mode.

We move now to analyze the nonlinear interface modes of the system
(i.e., $\gamma\neq 0$), looking for the low-order nonlinear modes
centered at either side of the interface between the lattices. For
a given value of $\beta$, we solve Eqs. (\ref{eq:2}) numerically
with the help of a straightforward extension of the
multidimensional Newton-Raphson method used earlier in our
analysis of one-dimensional waveguide arrays~\cite{OL_molina}. We
also compute the linear stability properties of each mode. Results
are displayed in Figs.~1 
\begin{figure}[t]
\noindent\includegraphics[scale=.35,angle=0]{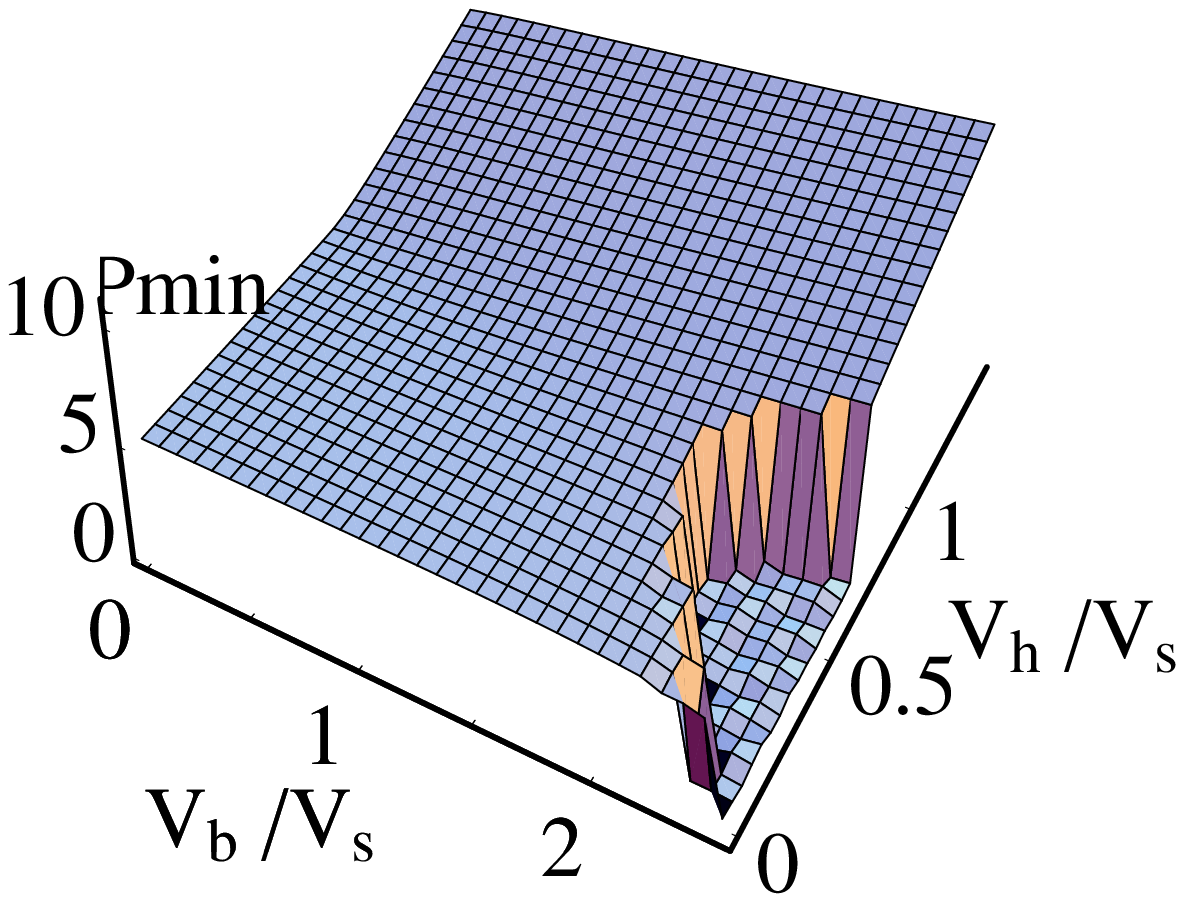}\hspace{0.0cm}
\includegraphics[scale=0.45,angle=0]{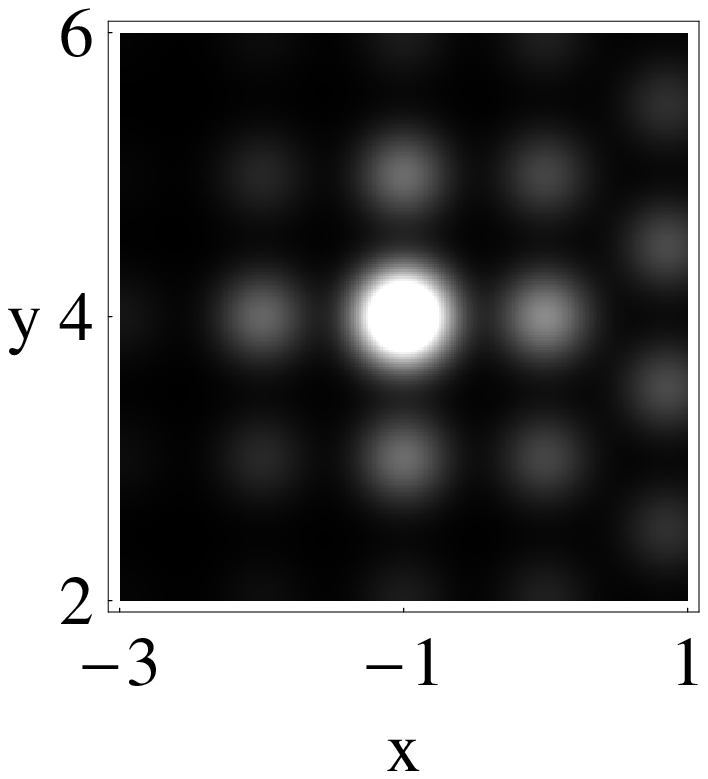}\hspace{6.0cm}
\vspace{0cm}
\includegraphics[scale=0.45,angle=0]{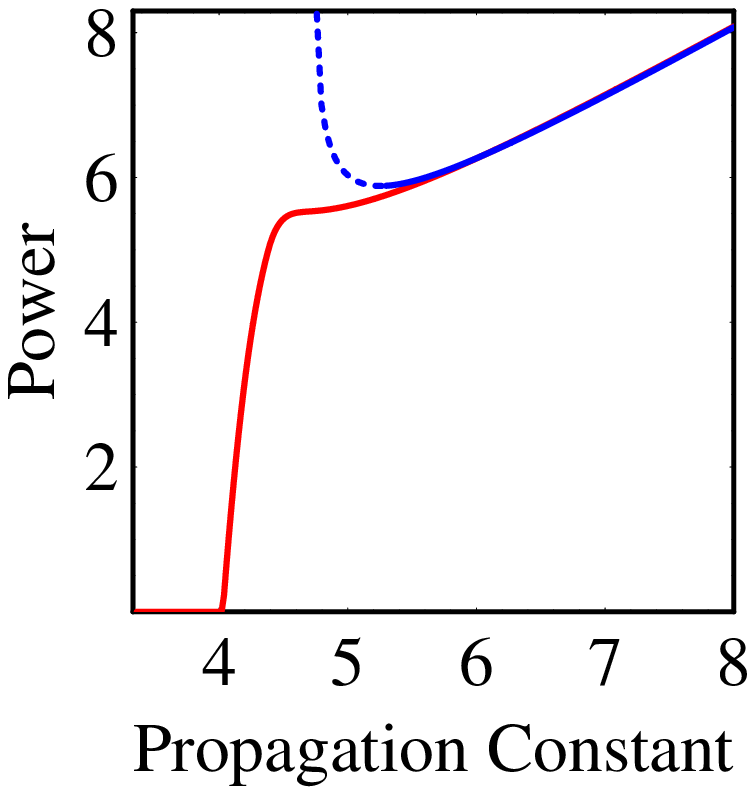}\hspace{0.0cm}
\includegraphics[scale=0.55,angle=0]{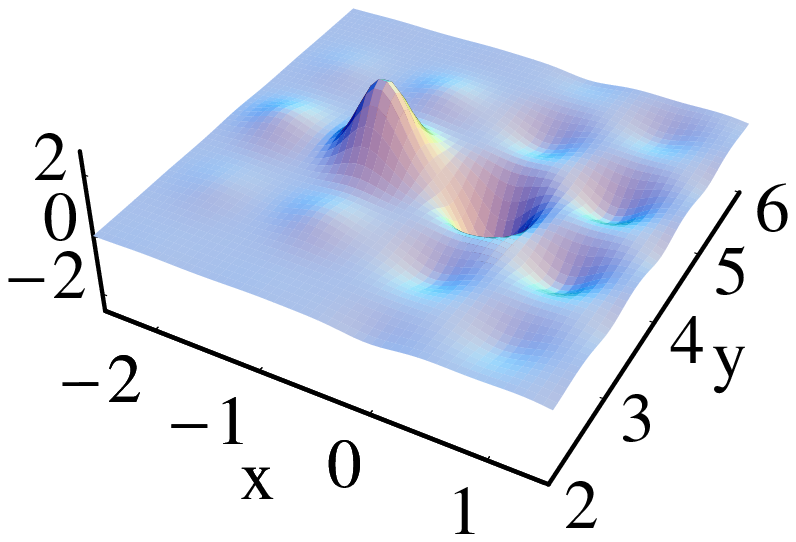}
\caption{(Color online) Nonlinear case ($\gamma>0$). Top left:
Minimum power to create a localized mode at the boundary of the
square lattice, as a function of the couplings. Top right: Example
of an interface localized mode centered on a site belonging to the
boundary of the square lattice ($V_{b}/V_{s}=1=V_{h}/V_{s},
\beta=5.9$). Bottom left: Power versus propagation constant for
interface modes localized at the boundary of the square lattice;
$V_{b}/V_{s}=1=V_{h}/V_{s}$ (red) and $V_{b}/V_{s}=0.8,
V_{h}/V_{s}=1$ (blue). The dashed portion of the curve denotes an
unstable regime. Bottom right: Example of a higher-order interface
mode.} \label{fig3}
\end{figure}
and 3, which show some examples of
low-order interface modes (Fig.~1) and the minimum power to effect
an interface mode at the square side of the boundary,
as a
function of the coupling parameters (Fig.~3).

The presence
\begin{figure}[t]
\includegraphics[scale=0.6,angle=0]{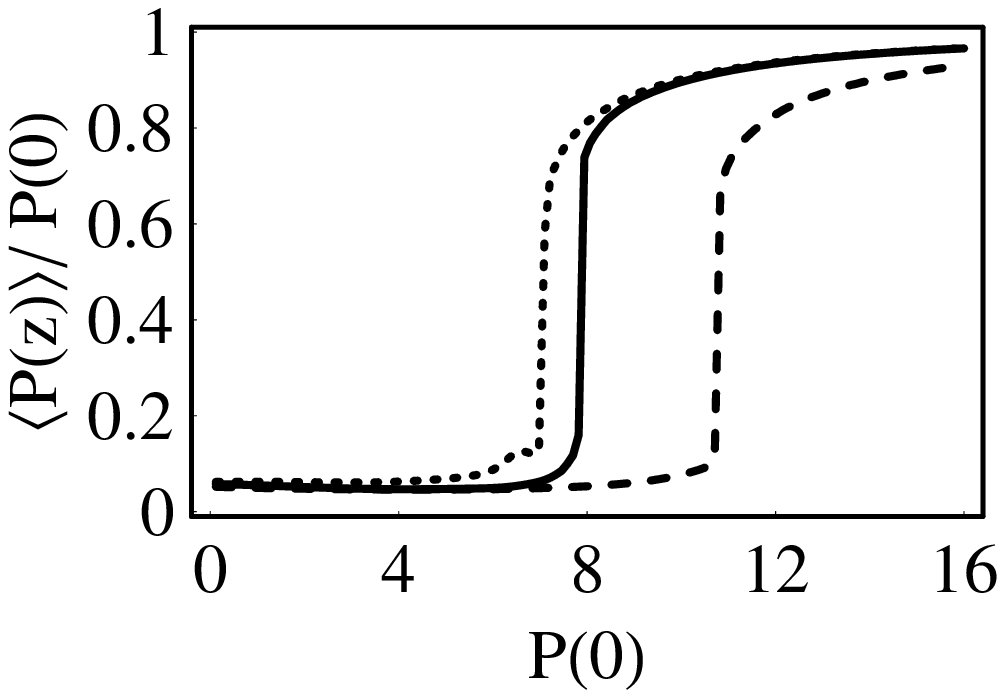}\\
\includegraphics[scale=0.5,angle=0]{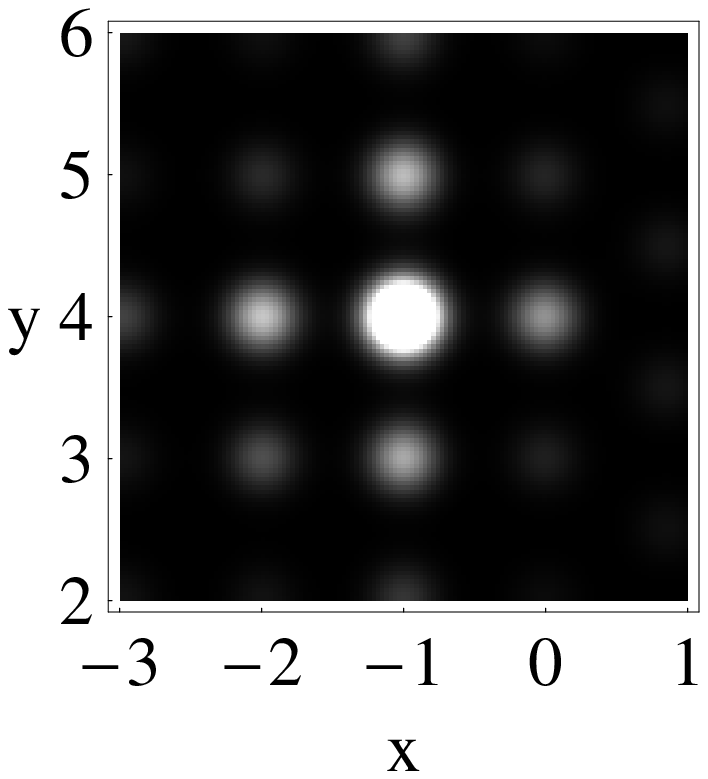}\hspace{0.cm}
\includegraphics[scale=0.5,angle=0]{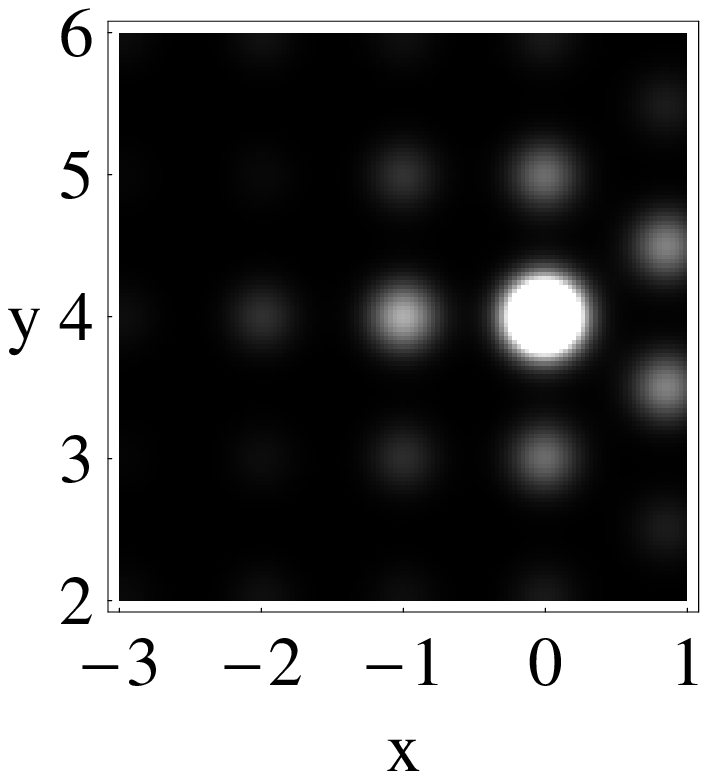}
\caption{Top: Average power fraction remaining at the
square-triangular interface waveguide, after some propagation
distance,  as a function of the input power (solid). The
dotted(dashed) curves refer to the cases of fully-square
(triangular) lattices. ($V_{b}/V_{s}=1= V_{t}/V_{s},
z_{max}=20/V_{s}$). Bottom: Dynamical excitation of an interface
localized mode. Left: localized mode at the boundary of the square
lattice. Right: Localized mode at the boundary of the triangular
lattice ($V_{b}/V_{s}=1, V_{t}/V_{s}=2/3$, input power: $2.6$).
Triangular lattice starts at $x=0$.} \label{fig4}
\end{figure}
of a steep ``hole'' in the power surface is due to the existence
of linear interface modes (Fig.~2).  The figure also shows an
example of such a mode, which is substantially narrower than its
linear counterpart. The stability analysis shows that all
nonlinear modes originating from linear ones are always stable,
while nonlinear modes outside the linear ``hole'' in the coupling
space require a minimum power to exist. For the latter, the
Vakhitov-Kolokolov stability criterium seems to hold. Finally, in
Fig.3 we also show for completeness, an example of a high-order
nonlinear mode, characterized by having positive amplitudes inside
the square sector, while inside the hexagonal lattice the
amplitudes are all negative. It could be described as a
``twisted'' mode along the direction perpendicular to the
boundary.

Finally, we examine the dynamical excitation of a localized
interface mode, excited by a highly-localized input beam launched
at a boundary waveguide. As expected, for given values of
$V_{b}/V_{s}$ and $V_{h}/V_{s}$, a narrow state is created above
some minimum power level. Figure 4 shows the average power
remaining at an initial waveguide located at the very boundary
between the square and triangular lattices
($V_{b}/V_{s}=1=V_{h}/V_{s}$), as a function of the input power.
Not surprisingly, the power curve falls between the curves for the
fully-square and fully-triangular lattices; with the fully-square
lattice possessing the smallest threshold power for self-trapping
due to its lower coordination number. Figure 4 also shows two
examples of dynamically generated interface modes.

In conclusion, we have studied localization of light at the
interface separating square and hexagonal photonic lattices and
determined the conditions for the existence of localized surface
states due to symmetry breaking. We have analyzed the effect of
the lattice topology and intersite couplings on the stability of
two-dimensional surface solitons which are found to differ
substantially from the one-dimensional discrete surface solitons.

This work was supported by Fondecyt grant 1080374 and by the
Australian Research Council.


\begin{thebibliography}{10}

\bibitem{makris_2D} K.G. Makris, J. Hudock, D.N. Christodoulides, G.
Stegeman, O. Manela, and M. Segev, Opt. Lett. {\bf 31}, 2774 (2006).

\bibitem{pla_our} R.A. Vicencio, S. Flach, M.I. Molina, and Yu.S.
Kivshar, Phys. Lett. A {\bf 364}, 274 (2007).

\bibitem{pre_2D} H. Susanto, P.G. Kevrekidis, B.A. Malomed, R.
Carretero-Gonz\'alez, and D.J. Franzeskakis, Phys. Rev. E {\bf 75}, 056605 (2007).

\bibitem{prl_1} X. Wang, A. Bezryadina, Z. Chen, K.G. Makris, D.N.
Christodoulides, and G.I. Stegeman, Phys. Rev. Lett. {\bf 98},
123903 (2007).

\bibitem{prl_2} A. Szameit, Y.V. Kartashov, F. Dreisow, T. Pertsch,
S. Nolte, A. T\"unnermann, and L. Torner, Phys. Rev. Lett. {\bf 98}, 173903 (2007).

\bibitem{ol_szameit} A. Szameit, Y. V. Kartashov, V.A. Vysloukh,
M. Heinrich, F. Dreisow, T. Pertsch, S. Nolte, A. T\"unnermann, F. Lederer,
and L. Torner, Opt. Lett. {\bf 33}, 1542 (2008).

\bibitem{OL_george} K.G. Makris, S. Suntsov, D.N. Christodoulides, G.I.
Stegeman, and A. Hach\'e, Opt. Lett. \textbf{30,} 2466 (2005).

\bibitem{PRL_george} S. Suntsov, K.G. Makris, D.N. Christodoulides, G.I.
Stegeman, A. Hach\'e, R. Morandotti, H. Yang, G. Salamo, and M.
Sorel, Phys. Rev. Lett. \textbf{96,} 063901 (2006).

\bibitem{OL_molina}
M. Molina, R. Vicencio, and Yu.~S. Kivshar,
Opt. Lett. {\bf 31}, 1693 (2006).

\bibitem{interface_2D} A. Szameit, Y.V. Kartashov, F. Dreisow, M. Heinrich,
V.A. Vysloukh, T. Pertsch, S. Nolte, A. T\"unnermann, F. Lederer, and L. Torner, Opt. Lett. {\bf 33}, 663 (2008).


\end{thebibliography}
\end{document}